\documentclass[aps,prl,10pt,twocolumn,showpacs]{revtex4-1}
\usepackage{graphicx,amsmath,time,mathrsfs}
\usepackage{hyperref}
\usepackage{times}

\newcommand{\cre}[1]{\hat{#1}^\dagger}
\newcommand{\des}[1]{\hat{#1}}
\newcommand{\therm}{\mathrm{th}}
\newcommand{\opt}{\mathrm{opt}}
\newcommand{\inp}{\mathrm{in}}
\newcommand{\ex}{\mathrm{ext}}

\hypersetup{
colorlinks=true,
linkcolor=blue,          
citecolor=blue,          
filecolor=magenta,       
urlcolor=cyan
}

\begin{document}

\title{Quantum limit of laser cooling in dispersively- and dissipatively-coupled optomechanical systems}

\author{Talitha~Weiss}
\affiliation{Department of Physics, University of Basel, Klingelbergstrasse 82, CH-4056 Basel, Switzerland}

\author{Andreas~Nunnenkamp}
\affiliation{Department of Physics, University of Basel, Klingelbergstrasse 82, CH-4056 Basel, Switzerland}

\date{\today}

\begin{abstract}
Mechanical oscillators can be cooled by coupling them to an optical or microwave cavity. Going beyond the standard quantum noise approach, we find an analytic expression for the steady-state phonon number in systems where the position of the mechanical oscillator modulates the cavity frequency as well as the cavity line width. We trace the origin for the quantum limit of cooling to fluctuations in the optical force both at and away from the mechanical frequency. Finally, we calculate the minimal phonon number for the different types of coupling. Our study elucidates how to beneficially combine dispersive and dissipative optomechanical coupling.
\end{abstract}

\pacs{42.50.-p, 07.10.Cm, 85.85.+j, 37.10.Vz}

\maketitle

\emph{Introduction.} Optomechanics is an area of research that is concerned with systems in which the position of a mechanical oscillator modulates the properties of an optical or microwave mode \cite{kv2008, mg2009, aghk2010, Meystre2013, Aspelmeyer2013}. Apart from fundamental questions, e.g.~about the decoherence of increasingly macroscopic objects \cite{Romero-Isart2010}, these systems have promising applications in information and quantum science, e.g.~transducers in quantum hybrid systems \cite{Stannigel2010}.

To enable these applications reducing the thermal motion of the mechanical oscillator has been a focus of intense research. Adapting laser-cooling techniques from atomic physics \cite{mccg2007,w-rnzk2007} experiments have observed the quantum ground state \cite{t2011, c2011, vdwsk2012} as well as an asymmetry in the mechanical sidebands \cite{s-nchakp2012,bbsbs-k2012}.

In Ref.~\cite{egc2009}, an optomechanical system has been introduced where the cavity line width depends parametrically on the position of a mechanical oscillator. A strikingly new  feature is a Fano line shape in the force spectrum that is a consequence of quantum noise interference. In a recent study we have shown that this form of force spectrum features two cooling and two instability regions \cite{wbn2013}. Following the proposal~\cite{xsh2011}, an interferometer setup has recently investigated these effects \cite{tkksh2012}.

For an optimal detuning between laser and cavity frequency the zero of the Fano line shape coincides with the mechanical frequency. As the quantum noise approach \cite{cdgms2010} estimates the fluctuations in the optical force with the noise spectrum at the mechanical frequency, it predicts the unphysical result that the phonon number goes to zero as the coupling is increased, i.e.~it does not give a quantum limit of cooling.

In this paper we go beyond this level of approximation and derive an analytic expression for the steady-state phonon number which takes into account the noise in the force spectrum at all frequencies. We show that noise away from the mechanical frequency can become the limiting process for cooling. In this case, the steady-state phonon number depends on the coupling in a qualitatively different way featuring a minimum at finite coupling. We give explicit expressions for the minimal phonon number for so-called purely dissipative as well as dissipative and dispersive coupling. While the Fano line shape due to dissipative coupling leads to a vanishing amplification rate, additional dispersive coupling can increase the cooling rate and, thus, further lower the phonon number. Our study provides the physical limit of cooling and shows how to exploit the presence of these two kinds of optomechanics.

\emph{Model.} We investigate an optomechanical system, where the resonance frequency $\omega_c$ of a cavity and its line width $\kappa$ are both modulated by the displacement of a mechanical oscillator with resonance frequency $\omega_M$. These two types of coupling between optical and mechanical degrees of freedom will be referred to as dispersive and dissipative coupling, respectively.

The Hamiltonian ($\hbar=1$) of such a system is given by $\hat{\mathscr{H}}=\omega_c\cre{a}\des{a}+\omega_M\cre{c}\des{c}+\hat{H}_\kappa+\hat{H}_\gamma+\hat{H}_\text{int}$, where $\cre{a}$($\des{a}$) are bosonic creation (annihilation) operators of the cavity mode, $\cre{c}$($\des{c}$) are bosonic creation (annihilation) operators of the mechanical mode, and $\hat{H}_\kappa$ and $\hat{H}_\gamma$ describe driving and damping of the cavity and the mechanical oscillator, respectively. The optomechanical coupling is given by \cite{egc2009}
\begin{equation}\label{Eq:Hint} 
\hat{H}_\text{int} = -\left[ \tilde{A}\kappa\hat{a}^\dagger\hat{a}+i\sqrt{\frac{\kappa}{2\pi\rho}}\frac{\tilde{B}}{2}\sum\limits_{q}^{}\left(\hat{a}^\dagger\hat{b}_q-\hat{b}_q^\dagger\hat{a}\right)\right]\frac{\hat{x}}{x_0},
\end{equation}
where $\tilde{A}\kappa = -\frac{d\omega_c(x)}{dx}x_0$ is the dispersive and $\tilde{B}\kappa =\frac{d\kappa(x)}{dx}x_0$ the dissipative coupling strength \footnote{For a detailed discussion of different possible dissipative couplings, see the Appendix.}. Here, $\hat{x}=x_0(\des{c}+\cre{c})$ denotes the displacement of the mechanical oscillator, $x_0=(2m\omega_M)^{-1/2}$ is the size of the zero-point fluctuations and $m$ is the mass of the mechanical oscillator, $\cre{b}_q$ ($\hat{b}_q$) are bosonic creation (annihilation) operators of the optical bath coupled to the cavity, and $\rho$ is the density of states of the optical bath that we treat as a constant for the relevant frequencies.

\begin{figure}
\centering
\includegraphics[width=0.8\columnwidth]{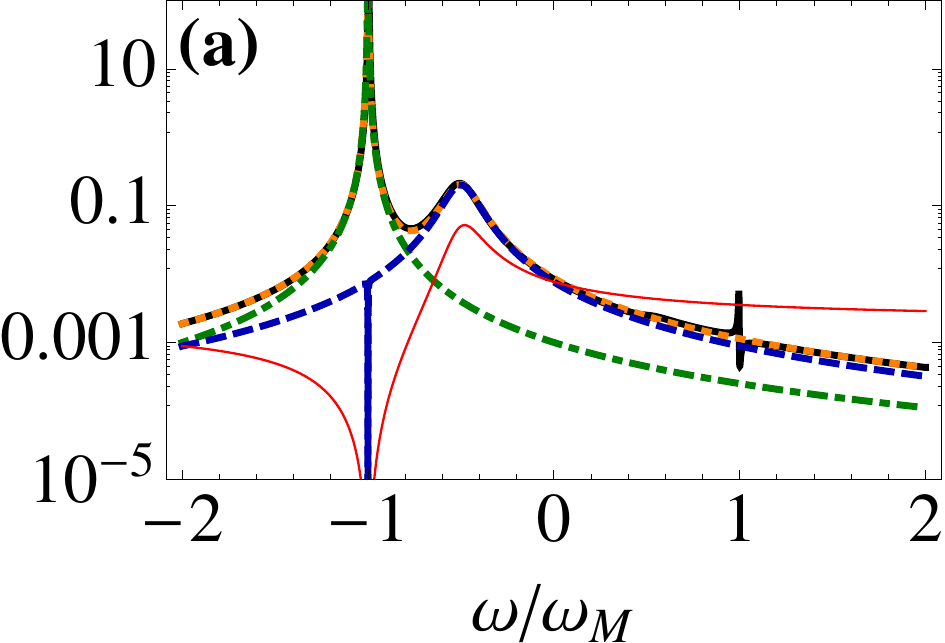}
\includegraphics[width=0.8\columnwidth]{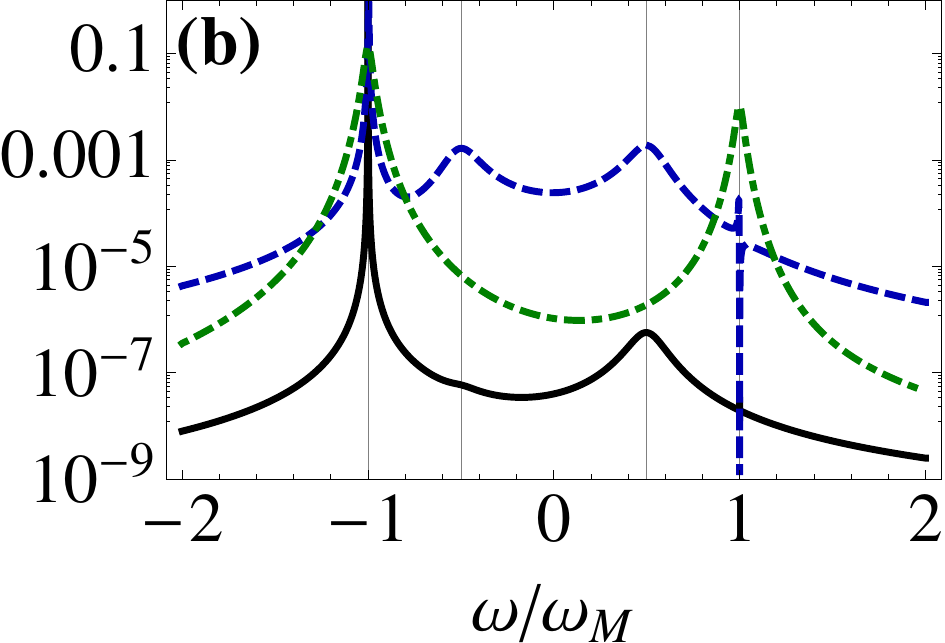}
\caption{(Color online) (a) The mechanical spectrum $S_{cc}(\omega)\omega_M$ obtained from the exact solution (solid black line) \cite{wbn2013} and the approximate expression (\ref{Eq:Scc}) (orange dotted line) for $\tilde{A}=0$, $\tilde{B}|\bar{a}|=0.2$, and $\Delta=\omega_M/2$. The green dot-dashed (blue dashed) line shows the first (second) term in Eq.~(\ref{Eq:Scc}). The thin red solid line is the weak-coupling force spectrum $S_{FF}(\omega)x_0^2/\omega_M$ (\ref{Eq:SFF}). (b) The optical output spectrum $S_{dd}^\text{out}(\omega)$ for $\tilde{A} = 0$, $\tilde{B}|\bar{a}|=0.01$, $\Delta=\omega_M/2$ (solid black line), $\tilde{A} = 0$, $\tilde{B}|\bar{a}|=0.2$, $\Delta=\omega_M/2$ (dashed blue line), and $\tilde{A}|\bar{a}| = 0.2$, $\tilde{B}=0$, $\Delta=-\omega_M$ (green dot-dashed line). Other parameters are $\omega_M/\gamma=10^{5}$, $n_\therm=100$, and $\omega_M/\kappa=5$.}
\label{Fig:Spec}
\end{figure}

We work in a frame rotating with the drive frequency $\omega_d$, using $\hat{a}=(\bar{a}+\hat{d})e^{-i\omega_dt}$, where $\bar{a}$ denotes the mean cavity amplitude and $\hat{d}$ denotes the fluctuations around this mean value. Then, using the input-output formalism \cite{cdgms2010} adapted for dissipatively coupled systems, we write down the linearized equations of motion \cite{egc2009,xsh2011,wm},
\begin{align}
\label{Eq:c}
\dot{\hat{c}} &= -\left(i \omega_M + \frac{\gamma}{2}\right) \hat{c} - \sqrt{\gamma} \hat{c}_\text{in}
+i x_0 \hat{F},\\
\dot{\hat{d}} &= \left(i\Delta-\frac{\kappa}{2}\right)\hat{d}-\sqrt{\kappa}\hat{d}_\mathrm{in}+\left[i\tilde{A}\kappa\bar{a}-\left(i\Delta+\frac{\kappa}{2}\right)\frac{\tilde{B}}{2}\bar{a}\right]\frac{\hat{x}}{x_0},\label{Eq:d}
\end{align}
with $\hat{F}x_0 = \tilde{A}\kappa\bar{a}^*\hat{d}+i\frac{\tilde{B}}{2}\bar{a}^*\sqrt{\kappa}\hat{d}_\mathrm{in}+i\frac{\tilde{B}}{2}\bar{a}^*(i\Delta+\frac{\kappa}{2})\hat{d}+h.c.$ Here, $\Delta=\omega_d-\omega_c$ is the detuning between the drive frequency $\omega_d$ and the cavity resonance $\omega_c$, and $\hat{d}_\text{in}$ ($\hat{c}_\text{in}$) describes the fluctuations in the optical (mechanical) input mode.
We assume Markovian baths, where the mechanical bath has a finite temperature $T$ and thus a thermal phonon number $n_\therm=(e^{\omega_M/k_B T}-1)^{-1}$ where $k_B$ denotes Boltzmann's constant, i.e.~$\langle\cre{c}_\text{in}(\omega)\des{c}_\text{in}(\omega')\rangle=2\pi\delta(\omega+\omega')n_\therm$ and $\langle\des{c}_\inp(\omega)\cre{c}_\inp(\omega')\rangle=2\pi\delta(\omega+\omega')(n_\therm+1)$, whereas, the optical bath is assumed to be at zero temperature, i.e.~$\langle\des{d}_\inp(\omega)\cre{d}_\inp(\omega')\rangle=2\pi\delta(\omega+\omega')$.

\emph{Noise contributions to the mechanical spectrum.} For weak coupling we can use the quantum noise approach to derive transition rates for the mechanical oscillator with Fermi's Golden Rule \cite{cdgms2010}. In our case, the force $\hat{F}$ leads to transitions between states with $n$ and $n \pm 1$ phonons. The rates are given by $\Gamma_{n\rightarrow n+1}=(n+1)\Gamma_\uparrow$ and $\Gamma_{n\rightarrow n-1}=n\Gamma_\downarrow$ with the amplification rate $\Gamma_\uparrow=x_0^2S_{FF}(-\omega_M)$ and cooling rate $\Gamma_\downarrow=x_0^2S_{FF}(\omega_M)$. They are obtained from the weak-coupling force spectrum $S_{FF}(\omega) = \int \! dt \, e^{i \omega t} \langle \cre{F}(t) \des{F}\rangle$ evaluated in the absence of coupling. Here, it is given by \cite{egc2009}
\begin{equation}\label{Eq:SFF}
S_{FF}(\omega) = \frac{\tilde{B}^2|\bar{a}|^2}{4x_0^2} \frac{\kappa (\omega+2\Delta-2\tilde{A}\kappa/\tilde{B})^2}{(\kappa/2)^2+(\omega+\Delta)^2}.
\end{equation}

As discussed in Ref.~\cite{egc2009}, for the optimal detuning $\Delta=\Delta_\text{opt}=\omega_M/2+\kappa\tilde{A}/\tilde{B}$ a Fano interference leads to a vanishing amplification rate $\Gamma_\uparrow=x_0^2S_{FF}(-\omega_M)=0$. The steady-state mean phonon number $n$ within the quantum noise approach is given by $n = (\gamma n_\therm + \Gamma_\uparrow)/(\gamma + \Gamma_\downarrow-\Gamma_\uparrow)$. If, however, $\Gamma_\uparrow = 0$ and $\Gamma_\downarrow \not= 0$, the mean phonon number $n$ goes to zero in the limit of large coupling strength, i.e., there is no quantum limit of cooling at this level of approximation.

In the following, we go beyond this standard quantum noise approach and take the complete force spectrum $S_{FF}(\omega)$ into account, i.e.~the noise at all frequencies. To do so, we solve Eq.~(\ref{Eq:d}) in the Fourier domain for $\hat{d}(\omega)$ and insert the result into the equation of motion of the mechanical oscillator (\ref{Eq:c}). Neglecting correlations between the optical field and the mechanical bath, e.g., $\langle \hat{d}(\omega) \hat{c}_\text{in}(\omega') \rangle = 0$, we obtain an approximation for the mechanical spectrum $S_{cc}(\omega) = \int \! dt \, e^{i \omega t} \langle \cre{c}(t) \des{c}\rangle$ as
\begin{equation}\label{Eq:Scc}
S_{cc}(\omega) = |\tilde{\chi}_M(-\omega)|^2 \left[ \gamma n_\therm + x_0^2 S_{FF}(\omega) \right],
\end{equation}
where $\tilde{\chi}_M(\omega) = [\tilde{\gamma}/2-i(\omega-\omega_M)]^{-1}$ is the effective mechanical response function, and where we have taken into account the optically-induced damping $\tilde{\gamma}=\gamma+\Gamma_\downarrow-\Gamma_\uparrow$ but have neglected the optically-induced frequency shift, i.e., the optical spring.

In Fig.~\ref{Fig:Spec} (a) we plot the mechanical spectrum $S_{cc}(\omega)$ for purely dissipative coupling $\tilde{A}=0$ and detuning $\Delta = \Delta_\text{opt}$. Within the quantum noise approach the mechanical spectrum has a peak at the mechanical frequency $\omega=-\omega_M$ describing the response of the mechanical oscillator to thermal fluctuations $\gamma n_\text{th}$ and optical force fluctuations at the mechanical frequency, i.e.~$S_{cc}(\omega)=|\tilde{\chi}_M(-\omega)|^2[\gamma n_\therm+x_0^2S_{FF}(-\omega_M)]$. In the resolved-sideband regime $\omega_M \gg \kappa$ the approximation (\ref{Eq:Scc}) features an additional peak due to optical force fluctuations $|\tilde{\chi}_M(-\omega)|^2  x_0^2 S_{FF}(\omega)$ missed by the quantum noise approach. For $\kappa\gg\tilde{\gamma}$, it is centered at $\omega = -\omega_M/2$ with a zero at $\omega = -\omega_M$ due to quantum noise interference \cite{egc2009}. In Fig.~\ref{Fig:Spec} (a), we plot the mechanical spectrum $S_{cc}(\omega)$ obtained from the exact solution to Eqs.~(\ref{Eq:c}) and (\ref{Eq:d}) as given in Ref.~\cite{wbn2013}. The agreement with the approximate expression (\ref{Eq:Scc}) is excellent.

The reason for the failure of the quantum noise approach can be understood by looking more closely at the force spectrum $S_{FF}(\omega)$ also in Fig.~\ref{Fig:Spec} (a). Within the quantum noise approach we approximate the optical force fluctuations by evaluating the force spectrum $S_{FF}(\omega)$ at the mechanical frequency. However, this is only justified if $S_{FF}(\omega)$ varies slowly around $\pm\omega_M$ on a scale of $\tilde{\gamma}$. For $\tilde{A}=0$ and $\Delta = \Delta_\text{opt}$ this is clearly not the case, and the quantum noise approach fails.

The two contributions to the mechanical spectrum $S_{cc}(\omega)$ in Eq.~(\ref{Eq:Scc}) can also be detected in the optical output spectrum $S_{dd}^\text{out}(\omega)=\int dt \, e^{i\omega t}\langle\cre{d}_\text{out}(t)\des{d}_\text{out}\rangle$ where the input-output relation is $\hat{d}_\mathrm{in}-\hat{d}_\mathrm{out}=-\sqrt{\kappa}\hat{d}-\sqrt{\kappa}\tilde{B}\bar{a}\hat{x}/2x_0$ \cite{wbn2013}. In Fig.~\ref{Fig:Spec} (b) we show the optical output spectrum $S_{dd}^\text{out}(\omega)$ for detuning $\Delta=\omega_M/2$. It features a dominant peak of width $\tilde{\gamma}$ at $\omega=-\omega_M$, a sharp dip to zero at $\omega=+\omega_M$, and two smaller peaks of width $\kappa$ at $\omega = \pm\omega_M/2$. The one at $\omega=+\omega_M/2$ exists at small coupling, whereas, the one at $\omega=-\omega_M/2$ appears at a larger coupling strength. In the limit of strong coupling the two peaks have equal weight. This differs significantly from the optical output spectrum $S_{dd}^\text{out}(\omega)$ for purely dispersive coupling $\tilde{B} = 0$ and detuning $\Delta = -\omega_M$ which features the well-known mechanical sidebands, i.e., two peaks of width $\tilde{\gamma}$ at $\omega=\pm \omega_M$, symmetric around the drive frequency. The different optical output spectra $S_{dd}^\text{out}(\omega)$ are a signature of the different cooling processes for these two kinds of couplings.

\emph{Improved expression for the mean phonon number.} Given the improved approximation for the mechanical spectrum $S_{cc}(\omega)$, Eq.~(\ref{Eq:Scc}), we obtain an expression for the mean phonon number $n$ by integrating over all frequencies $\omega$, i.e., $n=\langle \cre{c} \des{c} \rangle = \int S_{cc}(\omega) d\omega/(2\pi)$ where we have the analytic result,
\begin{eqnarray}\label{Eq:n}
n &=& \frac{\gamma n_\therm}{\tilde{\gamma}}+ \frac{\tilde{B}^2 |\bar{a}|^2}{4} \frac{ \kappa (-\omega_M + 2 \Delta - 2\tilde{A} \kappa/\tilde{B})^2}{\tilde{\gamma} [(\tilde{\gamma} + \kappa)^2/4 +   (\Delta - \omega_M)^2]} \\
&&+\frac{\tilde{B}^2 |\bar{a}|^2}{4}\frac{\tilde{\gamma} \kappa + 4 \Delta^2 - 16 \tilde{A} \Delta \kappa/\tilde{B} + (1 + 16 \tilde{A}^2/\tilde{B}^2) \kappa^2}{(\tilde{\gamma} + \kappa)^2 +  4 (\Delta - \omega_M)^2}.\nonumber
\end{eqnarray}
The first term in Eq.~(\ref{Eq:n}) accounts for the thermal fluctuations due to the mechanical bath with thermal phonon number $n_\therm$ reduced by the optically-induced damping. In the limit $\kappa\gg \tilde{\gamma}$, the second term simplifies to $x_0^2 S_{FF}(-\omega_M)/\tilde{\gamma}$, i.e., together with the first term it gives the standard quantum noise result. Note that the second term vanishes at the detuning $\Delta=\Delta_\text{opt}$ and thus, it does not provide a quantum limit of cooling. The third term in Eq.~(\ref{Eq:n}) goes beyond the quantum noise approach and is non-zero at the optimal detuning $\Delta = \Delta_\text{opt}$, thus, it leads to a quantum limit of cooling. For purely dissipative coupling $\tilde{A}=0$, Eq.~(\ref{Eq:n}) coincides with an expression in Ref.~\cite{xsh2011}. 

\begin{figure}
\centering
\includegraphics[width=0.8\columnwidth]{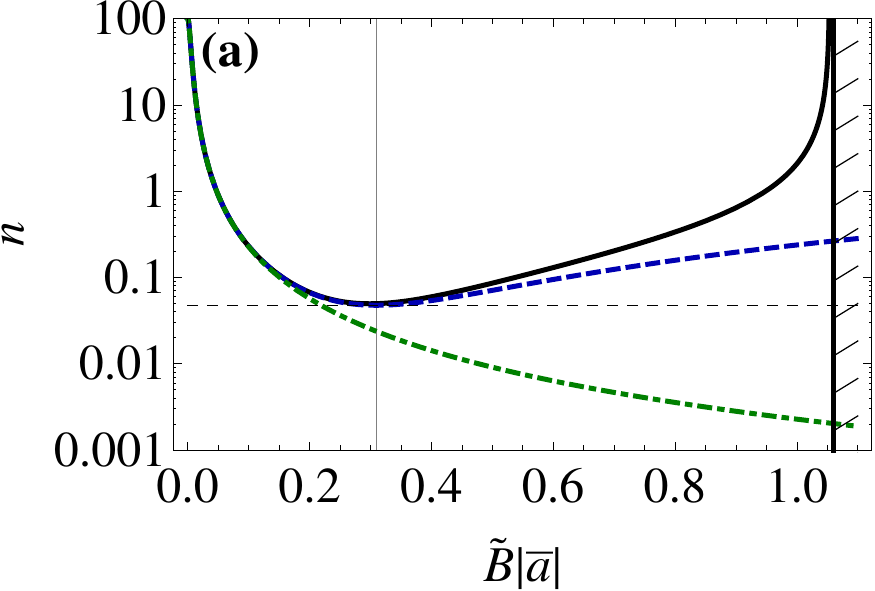}
\includegraphics[width=0.8\columnwidth]{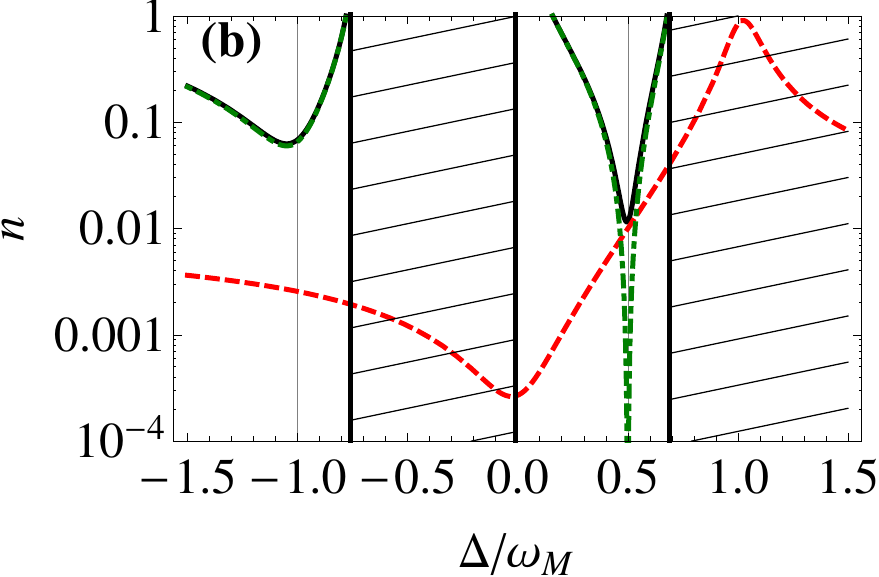}
\includegraphics[width=0.8\columnwidth]{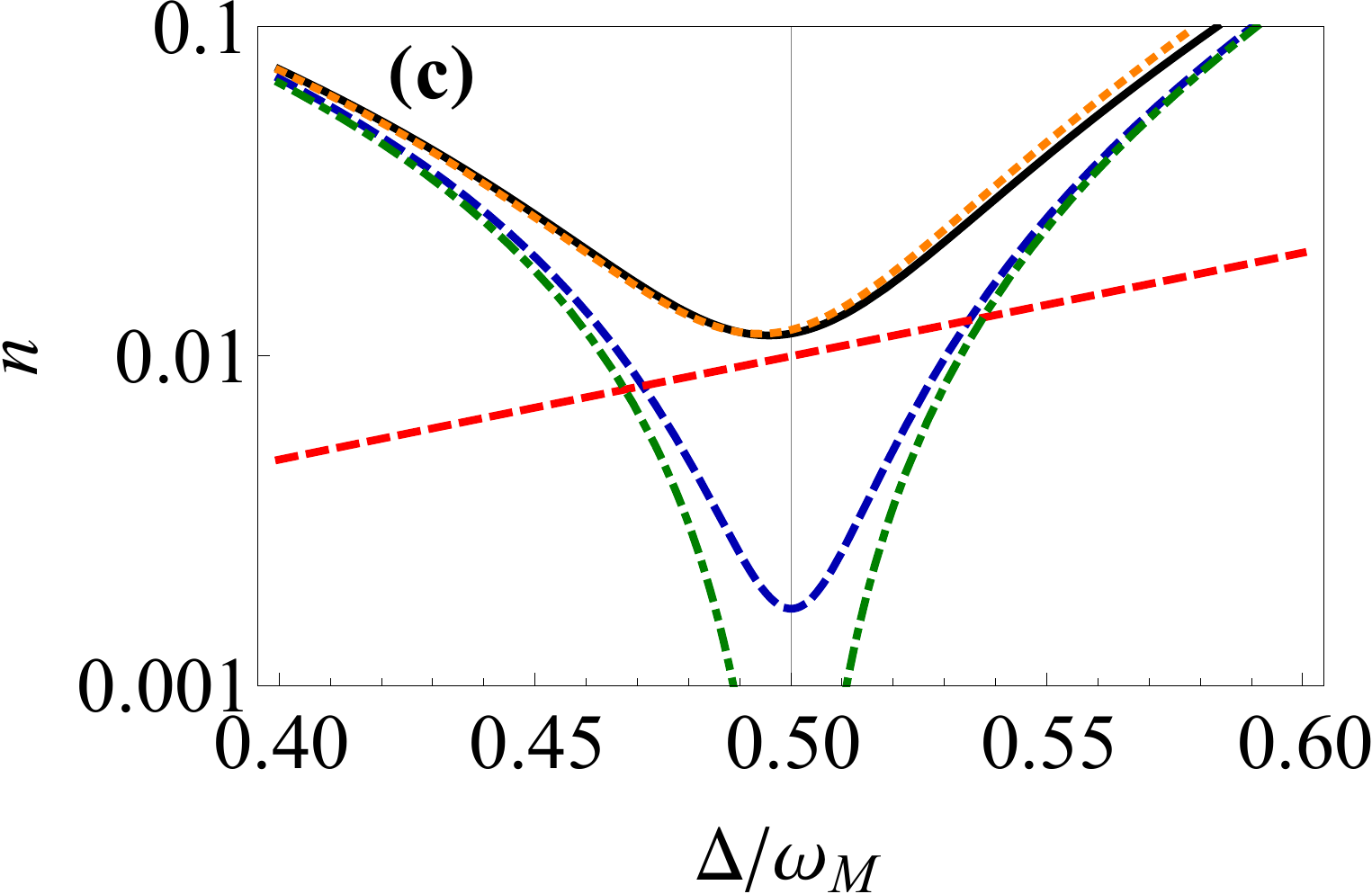}
\caption{(Color online) (a) Mean phonon number $n$ obtained from the quantum noise approach (green dot-dashed line), from Eq.~(\ref{Eq:n}) (blue dashed line), and the exact solution (solid black line) for $\tilde{A}=0$, $\omega_M/\kappa=3$, $\Delta =\omega_M/2$, $n_\therm=100$, and $\omega_M/\gamma=3 \cdot 10^{5}$. (b) Mean phonon number $n$ from Eq.~(\ref{Eq:n}) as a function of detuning $\Delta$ (solid black line). The green dot-dashed (red dashed) line shows the second (third) term of Eq.~(\ref{Eq:n}). Parameters are $\tilde{A}=0$, $\omega_M/\kappa=3$, $\tilde{B}|\bar{a}|=0.2$, $n_\therm=100$, and $\omega_M/\gamma=10^7$. In (c) we show a close up of (b) at $\Delta\approx\omega_M/2$ and include the quantum noise result (blue dashed line) and the exact solution (orange dotted line). Hatched areas indicate unstable regions.}
\label{Fig:Phonons}
\end{figure}

For purely dissipative coupling $\tilde{A} = 0$ at the detuning $\Delta = \omega_M/2$ and in the limit $\kappa\gg\tilde{\gamma}$, Eq.~(\ref{Eq:n}) simplifies to
\begin{equation}\label{Eq:cc}
n = \frac{\gamma n_\therm}{\tilde{\gamma}} + \frac{\tilde{B}^2 |\bar{a}|^2}{4}.
\end{equation}
We see that the phonon number $n$ has a qualitatively different dependence on the coupling strength as compared to the purely dispersive coupling $\tilde{B}=0$ discussed above. It is surprising that the second term in Eq.~(\ref{Eq:cc}) depends only on the coupling strength $\tilde{B}|\bar{a}|$ and not on the sideband parameter $\omega_M/\kappa$.

In Fig.~\ref{Fig:Phonons} (a), we plot the mean phonon number $n$ as a function of the coupling strength $\tilde{B}|\bar{a}|$ for purely dissipative coupling $\tilde{A}=0$.  Within the quantum noise approach, the phonon number $n$ approaches zero in the limit $\tilde{B}|\bar{a}|\rightarrow\infty$. For small coupling, the approximation (\ref{Eq:n}) agrees well with the quantum noise approach but, in contrast to the quantum noise approach, it features a minimum at finite coupling which defines a finite minimal phonon number $n_\text{min}$. Following, e.g., Ref.~\cite{vgfbtgvza2007}, an exact but cumbersome and not very illuminating expression for the phonon number $n$ can be derived. It agrees very well with our approximate result (\ref{Eq:n}) and deviates from it only for large coupling strengths close to the unstable region.

In Fig.~\ref{Fig:Phonons} (b), we plot the mean phonon number $n$ given by Eq.~(\ref{Eq:n}) as a function of detuning $\Delta$. We find that there are two cooling regions close to $\Delta=-\omega_M$ and around $\Delta=\omega_M/2$ where the phonon number is significantly reduced relative to the thermal phonon number $n_\therm$, and two unstable regions \cite{wbn2013}. The contributions of the second and third terms in Eq.~(\ref{Eq:n}) are shown separately. Notably, the second term in Eq.~(\ref{Eq:n}) is the limiting term for the phonon number $n$ at most detunings $\Delta$, except close to $\Delta=\Delta_\text{opt}$ where it goes to zero. In this case, the third term in Eq.~(\ref{Eq:n}) becomes important and provides the quantum limit of cooling, whereas it can be safely neglected elsewhere. Figure \ref{Fig:Phonons} (c) is focusing on $\Delta\approx\Delta_\text{opt}$. It shows that our approximate expression (\ref{Eq:n}) leads to a much better agreement with the exact solution than the quantum noise approach. In the following, we will focus on the case of $\Delta=\Delta_\opt$ where deviations from the quantum noise result are most significant.

\emph{Minimal mean phonon number.} For the purely dispersive coupling $\tilde{B} = 0$, in the limit $\kappa\gg\tilde{\gamma}$, we obtain the well-known result $n = \gamma n_\therm/\tilde{\gamma}+x_0^2 S_{FF}(-\omega_M)/\tilde{\gamma}=(\gamma n_\therm + \Gamma_\uparrow)/(\gamma + \Gamma_\downarrow-\Gamma_\uparrow)$ from Eq.~(\ref{Eq:n}). This leads to a minimal phonon number $n_\text{min} = \Gamma_\uparrow/(\Gamma_\downarrow-\Gamma_\uparrow) = \kappa^2/(16 \omega_M^2)$ for $\Delta \approx -\omega_M$ and $\tilde{A} \rightarrow \infty$. In this case, ground-state cooling is only possible in the resolved-sideband limit $\omega_M \gg \kappa$ \cite{mccg2007,w-rnzk2007}.

In the case of the purely dissipative coupling $\tilde{A}=0$, Eq.~(\ref{Eq:cc}) features a minimal phonon number, 
\begin{equation}\label{Eq:nmin}
n_\text{min} = \sqrt{\frac{\gamma n_\therm}{4\kappa} \left(\frac{\kappa^2}{\omega_M^2}+9\right)} - \frac{\gamma}{16 \kappa} \left(\frac{\kappa^2}{\omega_M^2}+9\right)
\end{equation}
at $\tilde{B}^2 |\bar{a}|^2 = \sqrt{\gamma n_\therm (\kappa^2/\omega_M^2+9)/\kappa}-\gamma(\kappa^2/\omega_M^2+9)/(4\kappa)$. Remarkably, in the good-cavity limit $\omega_M \gg \kappa$, $n_\text{min}$ becomes independent of the sideband parameter $\omega_M / \kappa$ and approaches a finite value $n_\text{min} = \sqrt{9\gamma n_\therm/(4\kappa) } - 9 \gamma/(16\kappa)$.

\begin{figure}
\centering
\includegraphics[width=0.8\columnwidth]{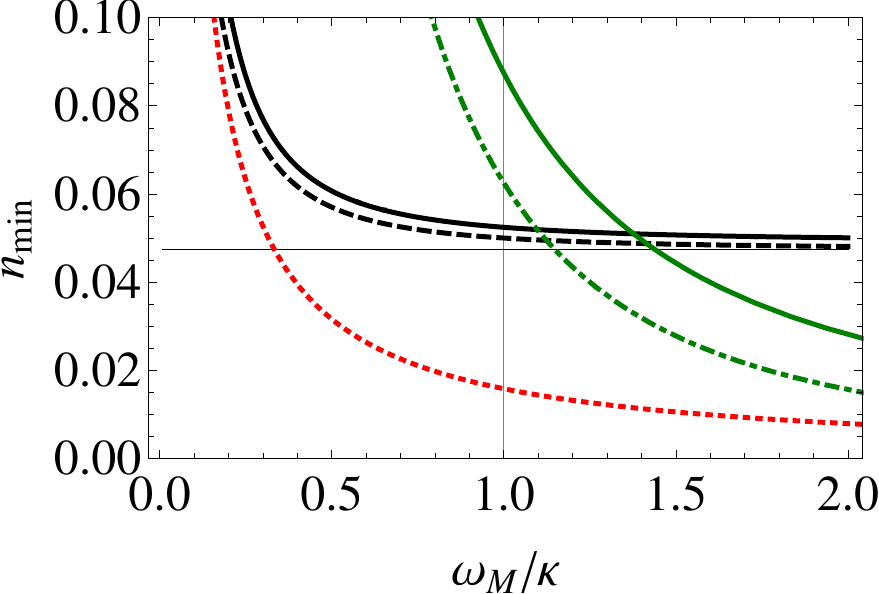}
\caption{(Color online) Minimal phonon number $n_\text{min}$ as a function of the sideband parameter $\omega_M/\kappa$ obtained from Eq.~(\ref{Eq:nmin}) (black dashed line) and the exact result (black solid line) for $\tilde{A}=0$ and $\Delta =\omega_M/2$; the thin solid black line indicates the limit of $n_{\min}$ for $\omega_M/\kappa\rightarrow\infty$. The dot-dashed green line gives the quantum noise result for purely dispersive coupling (at $\Delta=-\omega_M$, $\tilde{A}\rightarrow\infty$), and the green solid line shows the exact result, optimized for the dispersive coupling strength $\tilde{A}|\bar{a}|$ ($\tilde{B}=0$). The red dotted line shows the minimal phonon number for ideally mixed coupling at $\Delta=\omega_M/2+\kappa\tilde{A}/\tilde{B}$, Eq.~(\ref{Eq:nminAB}). Other parameters are $n_\text{th}=100$ and $\kappa/\gamma=10^{5}$.}
\label{Fig:Min}
\end{figure}

In systems featuring both dispersive and dissipative coupling, i.e., $\tilde{A}\neq 0$ and $\tilde{B}\neq 0$, at $\Delta=\Delta_\opt$ and in the limit $\kappa\gg\tilde{\gamma}$, Eq.~(\ref{Eq:n}) again simplifies to Eq.~(\ref{Eq:cc}). Remarkably, the second term of Eq.~(\ref{Eq:n}) is independent of the dispersive coupling strength $\tilde{A}$, and the minimal  phonon number $n_{\min}$ only depends on the coupling strength $\tilde{A}$ via the damping rate $\tilde{\gamma}$.

Minimizing Eq.~(\ref{Eq:cc}) over the coupling strength $\tilde{A}$ leads to $\tilde{A}\kappa = -3\tilde{B}\omega_M/2$. In this case, the optimal dissipative detuning $\Delta=\omega_M/2+\kappa\tilde{A}/\tilde{B}$ coincides with the optimal detuning for dispersive cooling, $\Delta=-\omega_M$. Within the quantum noise approach, this means taking advantage of a vanishing amplification rate $\Gamma_\uparrow=0$ at a detuning where dispersive coupling can increase the cooling rate $\Gamma_\downarrow$ and, thus, $\tilde{\gamma}$.
The minimal phonon number for this optimal mixed coupling is at $\tilde{B}^2|\bar{a}|^2 = \sqrt{n_\therm\gamma(\kappa^2/\omega_M^2)/\kappa} - \gamma (\kappa^2 /\omega_M^2)/(4 \kappa)$, and is
\begin{equation}\label{Eq:nminAB}
n_\text{min} = \sqrt{\frac{\gamma n_\therm}{4\kappa}\frac{\kappa^2}{\omega_M^2}}-\frac{\gamma}{16\kappa}\frac{\kappa^2}{\omega_M^2}.
\end{equation}
In contrast to the case of $\tilde{A}=0$, i.e., Eq.~(\ref{Eq:nmin}), the minimal phonon number $n_{\min}$ in Eq.~(\ref{Eq:nminAB}) vanishes in the limit $\omega_M\gg\kappa$.

Note that this requires dispersive coupling $\tilde{A}$ and dissipative coupling $\tilde{B}$ to have opposite signs. Whereas the sign of the coupling strength does not enter for purely dispersive or purely dissipative coupling, the relative sign matters in the presence of both couplings. According to our definitions of $\tilde{A}$ and $\tilde{B}$, opposite signs imply that the cavity resonance frequency $\omega_c$ and the cavity line width $\kappa$ both increase (or decrease) with increasing displacement of the mechanical oscillator $x$. 

In Fig.~\ref{Fig:Min}, we plot the minimal phonon number $n_\text{min}$ for dispersive, dissipative, and mixed coupling as a function of the sideband parameter $\omega_M/\kappa$. We see that, in the case of purely dissipative coupling $\tilde{A} = 0$, the minimal phonon number $n_\text{min}$ approaches a finite value for $\omega_M \gg \kappa$. In contrast, $n_\text{min}$ for purely dispersive coupling $\tilde{B} = 0$ goes to zero in the limit $\omega_M/\kappa \rightarrow \infty$. We also show the exact solution and find that our approximations capture the qualitative behavior very well. Figure \ref{Fig:Min} also shows that ideally mixed coupling can overcome the cooling limit for systems with purely dissipative coupling.

In closing, we note that opposite signs of dispersive and dissipative coupling will also allow for cooling on-resonance. If $\tilde{A}\kappa=-\tilde{B}\omega_M/2$, the optimal detuning simplifies to $\Delta_\text{opt} =\omega_M/2+\kappa\tilde{A}/\tilde{B}=0$. This would be particularly beneficial in the resolved-sideband limit $\omega_M \gg \kappa$ where, off-resonance, most of the input power is reflected off the cavity.

\emph{Acknowledgements.} We thank C.~Bruder, F.~Marquardt, J.~Teissier, and P.~Treutlein's group for interesting discussions. This work was financially supported by the Swiss NSF and the NCCR Quantum Science and Technology.

\appendix

\section{Appendix on different dissipative couplings}
Dissipative coupling arises if the cavity line width $\kappa$ is modulated by the mechanical displacement $x$, i.e., $\kappa = \kappa(x)$. In the main text, we have discussed the maximally overcoupled case where there is only a single loss channel for photons, so all losses are due to the channel where the coherent drive enters.

In general, several loss channels for photons can be present in an optomechanical system. We write the total cavity damping $\kappa=\kappa_\mathrm{ext}+\kappa_0$ as the sum of an external loss rate $\kappa_\mathrm{ext}$ associated with the channel where the drive enters and $\kappa_0$ that contains the losses through all other channels. 
Including the additional loss channel leads to an additional input mode $\hat{a}_\inp^0$ distinct from the input mode $\hat{a}_\inp^\ex$ associated with the drive, i.e., $\langle\hat{a}_\inp^0\rangle=0$ but $\langle\hat{a}_\inp^\mathrm{ext}\rangle=\bar{a}_\inp$. It is this difference that finally results in different expressions for the force $\hat{F}$ below. 

It is crucial to distinguish between the case,
\begin{equation}
\label{case1}
\kappa=\kappa_\mathrm{ext}(x)+\kappa_0,
\end{equation}
where the cavity line width associated with the drive port is modulated by the displacement and the case
\begin{equation}
\label{case2}
\kappa=\kappa_\mathrm{ext}+\kappa_0(x),
\end{equation}
where the internal losses depend on the displacement. 

\emph{Case 1.} The first case (\ref{case1}) is a generalization of the treatment in the main text where we used $\kappa=\kappa_\ex(x)$, i.e., $\kappa_0=0$. If $\kappa_0\neq 0$, $\tilde{B}\kappa_\ex=\frac{d\kappa_\ex(x)}{dx}x_0$ leads to the force, 
\begin{align}\label{Eq:forceEx}
\hat{F}x_0&=\tilde{A}\kappa\left(\bar{a}^*\des{d}+\bar{a}\cre{d}\right)+i\tilde{B}\sqrt{\kappa_\ex}\left[\bar{a}^*\hat{d}_\inp^\ex-(\hat{d}_\inp^{\ex})^\dagger\bar{a}\right]\nonumber\\
&-\frac{\tilde{B}}{2}\left[-i\bar{a}^*\left(i\Delta+\frac{\kappa}{2}\right)\des{d}-i\bar{a}\left(i\Delta-\frac{\kappa}{2}\right)\cre{d}\right],
\end{align}
where $\hat{d}_\inp^{\ex}$, $(\hat{d}_\inp^{\ex})^\dagger$ are the fluctuations of the input mode and the second line of Eq.~(\ref{Eq:forceEx}) is due to the coherent drive entering through the same port. This leads to the force spectrum \cite{ecg2009E},
\begin{align}\label{Eq:specEx}
S_{FF}(\omega)&= \frac{\tilde{B}^2|\bar{a}|^2}{4x_0^2}\frac{\kappa_\ex(\omega+2\Delta-2\tilde{A}\kappa/\tilde{B})^2}{(\kappa/2)^2+(\omega+\Delta)^2}\nonumber\\
&+\frac{\tilde{B}^2|\bar{a}|^2}{4x_0^2}\frac{\kappa_0[(\Delta-2\tilde{A}\kappa/\tilde{B})^2+\kappa^2/4]}{(\kappa/2)^2+(\omega+\Delta)^2}.
\end{align}
As discussed in Ref.~\cite{ecg2009E}, the optimal detuning $\Delta_\opt$ no longer leads to an exact zero of the force spectrum $S_{FF}(\omega)$ due to the second term of Eq.~(\ref{Eq:specEx}). Depending on the ratio of $\kappa_\ex$ and $\kappa_0$, the quantum noise interference becomes less perfect, and ultimately, if $\kappa_0\gg\kappa_\ex$, the force spectrum is a Lorentzian.

\emph{Case 2.} In the second case (\ref{case2}), where $\kappa=\kappa_\ex+\kappa_0(x)$, the force differs significantly from Eq.~(\ref{Eq:forceEx}). Since the input mode associated with $\kappa_0(x)$ describes only fluctuations of a zero-temperature bath, the force is given by
\begin{equation}
\hat{F}x_0=\tilde{A}\kappa\left(\bar{a}^*\des{d}+\bar{a}\cre{d}\right)+i\tilde{B}\sqrt{\kappa_0}\left[\bar{a}^*\hat{d}_\inp^0-(\hat{d}_\inp^{0})^\dagger\bar{a}\right].
\end{equation}
Notably, a term corresponding to the second line of Eq.~(\ref{Eq:forceEx}) is missing since this loss channel is not associated with a drive. Thus, for the purely dissipative coupling $\tilde{A}=0$, there is only one contribution to the force and no quantum noise interference. The force spectrum simplifies to $S_{FF}(\omega)=\kappa_0 \tilde{B}^2|\bar{a}|^2/(4x_0^2)$, i.e., it becomes completely flat as a function of frequency.

In the presence of both types of coupling, the dispersive coupling term provides a cavity-mediated force, whereas, dissipative coupling leads to a force directly proportional to the optical bath mode $\hat{d}_\inp^0$, and these two contributions can interfere. Using $\tilde{B}\kappa_0=\frac{d\kappa_0(x)}{dx}x_0$, the force spectrum is given by
\begin{align}
S_{FF}(\omega)&= \frac{\tilde{B}^2|\bar{a}|^2}{4x_0^2}\frac{\kappa_0[(\omega+\Delta-2\tilde{A}\kappa/\tilde{B})^2+\kappa^2/4]}{(\kappa/2)^2 + (\omega+\Delta)^2}\nonumber\\
&+\frac{\tilde{B}^2|\bar{a}|^2}{4x_0^2}\frac{\kappa_\ex(2\tilde{A}\kappa/\tilde{B})^2}{(\kappa/2)^2+(\omega+\Delta)^2},
\end{align}
i.e., it has a Fano line shape that, in contrast to the first case, becomes a Lorentzian if $\kappa_\ex\gg\kappa_0$. Also in contrast to the first case, no perfect destructive interference is possible, i.e.~the force spectrum $S_{FF}(\omega)$ has no exact zero. The optimal detuning from the first case no longer has a special meaning in this context and does not lead to a vanishing term of $S_{FF}(-\omega_M)$. 

Finally, note that the relative sign of dispersive and dissipative coupling, i.e., of $\tilde{A}$ and $\tilde{B}$, becomes relevant again.

\end{document}